\documentclass[12pt]{article}
\input epsf
\usepackage{amssymb}
\usepackage{amsfonts}
\usepackage{psfrag}
\usepackage{graphicx}

\renewcommand{\a}{\alpha}

\newcommand{\rmd}{{\rm d}}
\newcommand{\m}{\mu}
\newcommand{\n}{\nu}

\def\be{\begin{equation}}
\def\ee{\end{equation}}
\def\bea{\begin{eqnarray}}
\def\eea{\end{eqnarray}}
\def\ba{\begin{array}}
\def\ea{\end{array}}
\def\bi{\begin{itemize}}
\def\ei{\end{itemize}}

\catcode`\@=11
\def\@citex[#1]#2{%
\if@filesw \immediate \write \@auxout {\string \citation {#2}}\fi
\@tempcntb\m@ne \let\@h@ld\relax \def\@citea{}%
\@cite{%
  \@for \@citeb:=#2\do {%
    \@ifundefined {b@\@citeb}%
      {\@h@ld\@citea\@tempcntb\m@ne{\bf ?}%
      \@warning {Citation `\@citeb ' on page \thepage \space undefined}}%
      {\@tempcnta\@tempcntb \advance\@tempcnta\@ne%
      \@tempcntb\number\csname b@\@citeb \endcsname \relax%
      \ifnum\@tempcnta=\@tempcntb %Number follows previous--hold on to it
        \ifx\@h@ld\relax%
          \edef \@h@ld{\@citea\csname b@\@citeb\endcsname}%
        \else%
          \edef\@h@ld{\ifmmode{-}\else--\fi\csname b@\@citeb\endcsname}%
        \fi%
      \else%   %  non-successor--dump what's held and do this one
        \@h@ld\@citea\csname b@\@citeb \endcsname%
        \let\@h@ld\relax%
      \fi}%
    \def\@citea{,\penalty\@highpenalty\,}%
  }\@h@ld
}{#1}}

\def\@citeb#1#2{{[#1]\if@tempswa , #2\fi}}
\def\@citeu#1#2{{$^{#1}$\if@tempswa , #2\fi }}
\def\@citep#1#2{{#1\if@tempswa , #2\fi}}

\def\bcites{         % cite with []'s
        \catcode`\@=11
        \let\@cite=\@citeb
        \catcode`\@=12
}

\def\upcites{         % cite with exponents
        \catcode`\@=11
        \let\@cite=\@citeu
        \catcode`\@=12
}

\def\plaincites{      % cite without brackets
        \catcode`\@=11
        \let\@cite=\@citep
        \catcode`\@=12
}

\newcount\hour
\newcount\minute
\newtoks\amorpm
\hour=\time\divide\hour by 60
\minute=\time{\multiply\hour by 60 \global\advance\minute by-\hour}
\edef\standardtime{{\ifnum\hour<12 \global\amorpm={am}%
        \else\global\amorpm={pm}\advance\hour by-12 \fi
        \ifnum\hour=0 \hour=12 \fi
        \number\hour:\ifnum\minute<10 0\fi\number\minute\the\amorpm}}
\edef\militarytime{\number\hour:\ifnum\minute<10 0\fi\number\minute}

\def\draftlabel#1{{\@bsphack\if@filesw {\let\thepage\relax
   \xdef\@gtempa{\write\@auxout{\string
      \newlabel{#1}{{\@currentlabel}{\thepage}}}}}\@gtempa
   \if@nobreak \ifvmode\nobreak\fi\fi\fi\@esphack}
        \gdef\@eqnlabel{#1}}
\def\@eqnlabel{}
\def\@vacuum{}
\def\marginnote#1{}
\def\draftmarginnote#1{\marginpar{\raggedright\scriptsize\tt#1}}
\def\draft{
        \pagestyle{plain}
        \overfullrule=2pt
        \oddsidemargin -.5truein
        \def\@oddhead{\sl \phantom{\today\quad\militarytime} \hfil
        \smash{\Large\sl DRAFT} \hfil \today\quad\militarytime}
        \let\@evenhead\@oddhead
        \let\label=\draftlabel
        \let\marginnote=\draftmarginnote
        \def\ps@empty{\let\@mkboth\@gobbletwo
        \def\@oddfoot{\hfil \smash{\Large\sl DRAFT} \hfil}
        \let\@evenfoot\@oddhead}
        \def\@eqnnum{(\theequation)\rlap{\kern\marginparsep\tt\@eqnlabel}%
        \global\let\@eqnlabel\@vacuum}  }
\begin{document}

\hfill CERN-TH/2002-021

\hfill {\tt hep-th/0302063} 
\vspace{ 0.2cm}

\begin{center}
\huge
{ \bf Evaluating the AdS dual of the critical 
$O(N)$ vector model} 
\normalsize

\vspace{2cm}
\large
{\bf
Anastasios C. Petkou}
\footnote{tassos.petkou@cern.ch} \\
\normalsize
\vspace{.2cm}
{\it 
CERN Theory Division, \\
CH-1211 Geneva 23, \\
Switzerland}
\vspace{.5cm}

\end{center}

\vspace{0.8cm}
\large
\centerline{\bf Abstract}
\normalsize
\vspace{.8cm}

We argue that the AdS dual of the three dimensional critical $O(N)$
vector model can be evaluated using the Legendre transform that relates 
the generating 
functionals of the free UV and the interacting IR fixed points of the
boundary theory. As an example, we use our proposal to evaluate
the minimal bulk action of
the scalar field that it is dual to the spin-zero ``current'' of the
$O(N)$ vector model. We  find that
the cubic bulk self interaction  
coupling vanishes. We briefly discuss the implications of our results
for higher 
spin theories and comment on the bulk-boundary duality
for subleading $N$.    
 
\newpage

\section{Introduction}

The relation between gauge fields and strings has been significantly
enlightened by the AdS/CFT correspondence
\cite{adscft}. The general picture emerging is that
the large tension 
limit of string theory corresponds, holographically, to strongly coupled gauge
theories. Nevertheless, one would
ideally like to go further and understand the stringy picture
of weakly coupled 
gauge theories. The small tension limit of
string theory is an obvious candidate for this picture and a semiclassical
description of it would be desirable. Higher spin gauge
theories \cite{HSVasiliev} might provide such a semiclassical 
description \cite{HMS}. Moreover, the formulation of higher spin
theories in AdS spaces \cite{FV} opens the possibility for an holographic
interpretation of weakly coupled gauge theories. Recent work on higher
spin theories includes \cite{highspin}.

Recently, it was suggested that an interesting laboratory for studying
the relation between weakly coupled quantum field theories and higher spin
theories is provided by the well-known three dimensional critical
$O(N)$ vector model. The explicit proposal put forward in \cite{KP} is
that both the free UV and the interacting IR fixed point of the
$O(N)$ vector model are holographically described by the same AdS$_4$
higher spin 
theory. A manifestation of such a  degeneracy in the holographic
description is the fact that the UV and IR generating
functionals of the critical $O(N)$ model are related by a Legendre
transform for large $N$. This is one step further from the standard
cases of AdS/CFT
correspondence where the relation between the weakly and strongly
coupled boundary CFTs, even for large $N$,  is in general unknown. The apparent 
puzzle of having to describe both a free CFT (which does not have  anomalous
dimensions), as well as an interacting one, (which here has anomalous
dimension of $O(1/N)$),  by
the  same AdS theory was recently argued to be resolved by a Higgs
mechanism for gauge fields with spin $>$2 in 
AdS$_4$ \cite{GPZ}. This mechanism is at work only when the bulk scalar is
quantized with boundary conditions 
such that is corresponds to an operator of dimension 2+$O(1/N)$ in the
boundary. On the other hand, it is known \cite{tassosON1,tassosON2}
that for subleading $N$ the massless degrees of freedom coupled at the
UV and IR fixed points of the critical $O(N)$ vector model are
different, hence the relation between the 
corresponding UV and IR generating functionals is less clear.

The Lagrangian for
the AdS$_4$ higher spin theory is implicitly known through
complicated field equations \cite{HSVasiliev}. This may be reminiscent
of the standard 
situation with the IIB SUGRA that is dual to ${\cal N}=4$ SYM, however
there is an important difference: in the case at hand one knows
explicitly both the weakly as well as the strongly coupled regimes of
the boundary field theory. Therefore, one can work from bottom-to-top and evaluate the
bulk theory using the knowledge of the boundary CFT.
In this work we propose that the  evaluation of the bulk
AdS$_4$ theory dual to the critical $O(N)$ vector model can be done by
a self-consistency procedure based on the Legendre transform that
relates the generating functionals of the free UV and the interacting
IR fixed points of the $O(N)$ vector model. To illustrate our idea, we
consider here the minimal ansatz for the AdS$_4$ Lagrangian and work
out the tree level bulk couplings up to the quartic one. 
This is done by successively matching  the correlation
functions produced by the bulk AdS$_4$ Lagrangian to the corresponding
ones of the boundary CFT which can be explicitly calculated. Although
we do not consider the coupling of the bulk scalar to higher spin
fields in AdS$_4$, our result for the bulk cubic self interaction coupling can be carried
over to the higher spin Lagrangian. Our calculation of the bulk quartic
self interaction coupling may be useful both as its stands
i.e. for a possible pure gravity dual of the $O(N)$ vector model or as
an intermediate result in
future calculations of the higher spin dual of the model.

In Section 2 we review the degeneracy in
the holographic correspondence for scalar fields in AdS$_{D+1}$ with mass $m$ in
the range $-D^2/4<m^2< 1-D^2/4$, and its manifestation in terms of the
Legendre transform that relates the UV and the IR generating
functionals of the boundary theory. In Section 3 we apply our proposal
to evaluate the bulk action up to the the quartic scalar self interaction
term. In Section 4 we briefly discuss the implications of our results for higher
spin theories and comment on the nature of the bulk-boundary relation
for subleading $N$. The Appendix is reserved for
a compact presentation of the many technical details.

\section{The degeneracy in the holographic description and the
Legendre transform} 

It was noticed already in  the early days of AdS/CFT that there is a
potential ambiguity in the holographic description of a boundary
theory \cite{KW}. Let
$\phi(r,x)$ be a bulk scalar\footnote{We use
throughout the Euclidean version of the Poincar\'e patch of AdS$_{D+1}$
with $\rmd s^2 =(\rmd r^2+\rmd x^2)/r^2$ where $x^i=(x^1,..,x^D)$ and
we set the AdS 
radius to 1 such that the cosmological constant $\Lambda =
-D(D-1)/2$.} with mass $m$. Its asymptotic 
behavior near the 
boundary of AdS$_{D+1}$ is
\be
\label{bbehav}
\phi(r,x)|_{r\rightarrow 0} \approx r^{\Delta_-}\phi_0(x )
+r^{\Delta_+}A(x)\,,\,\,\, \Delta_{\pm} = \frac{D}{2} \pm
\nu\,,\,\,\, \nu=\frac{1}{2}\sqrt{D^2 +4m^2}\geq 0\,.
\ee
The functions
$\phi_0(x)$ and $A(x)$ are the two necessary boundary data to
determine the solution of the second-order bulk equation of
motion for $\phi(r,x)$. Quantizing then $\phi(r,x)$ with boundary
condition $A(x)=0$ ($\phi_0(x)=0$) would give the generating functional
of the boundary operator ${\cal O}(x)$ with dimension $\Delta_+$
($\tilde{{\cal O}}(x)$ with dimension $\Delta_-$). The above ambiguity
does not show up in most of the studied cases of AdS/CFT where the operator
$\tilde{{\cal O}}(x)$ has dimension below the unitarity bound
i.e. $\Delta_- < D/2-1$.  

Nevertheless, there exist important cases where both $\Delta_\pm$ are
above the unitarity bound. Then, the quantization ambiguity
is present even when the 
asymptotic behavior (\ref{bbehav}) of $\phi(r,x)$ is determined by
one arbitrary boundary data as when one 
requires that the bulk solution vanishes in the far interior
($r\rightarrow \infty$) of AdS. In such a case the two functions
appearing in (\ref{bbehav}) are related by
\be
\label{brel}
A(x) = C_{\Delta_+}\int\rmd^D
y\frac{1}{(x-y)^{2\Delta_+}}\phi_0(y)\,,\,\,\,\, C_{\Delta_+} = 
\frac{\Gamma(\Delta_+)}{\pi^{\frac{D}{2}}\Gamma(\n)}\,.
\ee
Then, the application of AdS/CFT correspondence yields {\it
either} a functional $W[\phi_0]$ of $\phi_0(x)$ {\it or} a functional
$J[A]$ of $A(x)$. The first generates correlation functions of ${\cal
O}(x)$ and the second of $\tilde{{\cal O}(x)}$. However, due to
(\ref{brel}) the two functionals are not independent but one is the
Legendre transform of the other as \cite{KW,MV}
\be
\label{Legendre}
W[\phi_0] +2\nu\int\rmd^D x\,\phi_0(x)A(x) = J[A]\,,\,\,\,\,\,\, \frac{\delta
W[\phi_0]}{ \delta\phi_0(x)} = -2\n A(x)\,.
\ee
An interesting observation regarding the relation between $W[\phi_0]$
and $J[A]$ was made in \cite{witten_mtraces} and was further
elaborated in 
\cite{GM,GK}. The interchange between the boundary conditions
$\phi_0(x)=0$ and $A(x)=0$ is induced by a ``double-trace''
deformation\footnote{This fact was also implicit in the OPE analysis
of multi-trace deformations in \cite{tassos_mtraces}.}. One way to see
this is to first choose the boundary 
condition $\phi_0(x)=0$, which would yield the theory for the operator
$\tilde{{\cal O}}(x)$ with dimension $\Delta_-$, and then perturb this
theory by 
$\frac{f}{2}\tilde{{\cal O}}^2(x)$. Noting then that $D/2> \Delta_- >
D/2-1$, this perturbation is relevant and for $f\rightarrow 
\infty$ it leads to a
possible IR fixed point of the $\tilde{{\cal O}}(x)$ theory, and at
the same time to the boundary condition $A(x)=0$. Therefore, $J[A]$
may be viewed as the generating functional of the UV fixed point CFT
while $W[\phi_0]$ as the one for the IR fixed point CFT, the two being
connected by an RG flow. 

An application of the above phenomenon can be found in
the recently discussed case of the AdS dual of the critical
three-dimensional $O(N)$ vector model. It has been suggested in
\cite{KP} that this  well-known three-dimensional CFT has a dual in
AdS$_4$ which might be a higher spin theory. For example, one would expect
that there exists an action for a massive scalar on AdS$_4$ that
yields the generating functional for the spin-zero
``current''\footnote{This name is used for the operator proportional
  to $\phi^a(x)\phi^a(x)$ where 
$\phi^a(x)$, $a=1,2,..,N$ are the elementary fields.} of the free UV 
$O(N)$ CFT. On the other hand, the generating functional of
the interacting IR fixed point of the model gives the correlation
functions for 
a composite operator of dimension 2+$O(1/N)$. Then, by the arguments
above, these two
generating functional should be related by a Legendre transform as in
(\ref{Legendre}). However, the IR generating functional
is the one directly obtained from the bulk AdS$_4$ Lagrangian by the
standard AdS/CFT correspondence and involves the parameters of the
bulk Lagrangian. Moreover, the knowledge of higher-pt
correlation functions in the IR CFT gives information about anomalous
dimensions of various fields in the theory which can be directly translated
to information about corrections to masses of the bulk fields. In the next
section we describe explicitly the use of the Legendre transform as a
self-consistency condition to evaluate the parameters of the AdS$_4$
Lagrangian dual to the spin zero ``current'' of the critical $O(N)$
model up to quartic order.

\section{Evaluation of the AdS dual of the critical $O(N)$ vector
model}

\subsection{The cubic bulk coupling}

The proposal of \cite{KP} for the AdS dual of the spin zero
``current'' of the critical $O(N)$
vector model is to consider a conformally coupled scalar on
AdS$_4$. In this Subsection  we perform the calculations for general $D$
and keep in 
mind that at the end we want to set $D=3$. The minimal 
gravity action
that could reproduce the correlation functions of the spin-zero
``current'' of the critical $O(N)$ vector model is 
\be
\label{adsaction}
S_{D+1} = \frac{1}{2\kappa^2_{D+1}}\int\rmd^{D+1} x\sqrt{g}\left[ -R+ 2\Lambda+
\frac{1}{2}g^{\m\n}\partial_\m\phi\partial_\n\phi 
-\frac{D^2-1}{8}\phi^2 +\frac{g_3}{3!}\phi^3 +\frac{g_4}{4!}\phi^4+...\right]\,.
\ee
The overall normalization of the action can be fixed by requiring that
the coefficient $C_T$ of the energy momentum 2-pt function following
from (\ref{adsaction}) coincides with the one of the $O(N)$ vector
model. The latter is completely determined 
by the overall normalization of the $O(N)$ vector model action and
does not depend on the  normalization of the scalar
fields. Following the general treatment of the $O(N)$ vector model in 
\cite{tassosON1} we have
\be
\label{CTON}
C_T^{O(N)} = N\frac{D}{(d-1)S^2_D}\,,\,\,\,\,\,
S_{D}=\frac{2\pi^{\frac{D}{2}}}{\Gamma\left(\frac{D}{2}\right)}\,. 
\ee
Then, from (\ref{CTON}) and using the general result of \cite{LT} for $C_T$ we obtain to leading 
order in $O(1/N)$ 
\be
\label{CT}
\frac{1}{2\kappa^2_{D+1}} = N\frac{\pi^D}{(D+1)\Gamma(D)}\frac{1}{S_{D}^3}\,,\,\,\,\,\,
\frac{1}{2\kappa^2_4} = \frac{N}{2^9}\,.
\ee

Next we concentrate on correlation function of the operators dual to
$\phi(r,x)$. We first use the ``regular'' boundary data
$\phi_0(x)$, we set $\Delta\equiv \Delta_+$ and perform for simplicity
the rescaling 
$\phi_0\rightarrow \left(2\kappa_{D+1}^2\right)^\frac{1}{2} \phi_0$ to obtain the correlation
functions of normalized \footnote{Operators whose 2-pt function is of
order $O(1)$.} operators
\be
\label{Wphi0}
W[\phi_0] = \sum_{n=2}^{\infty}\frac{1}{n!}\int\rmd^D x_1...\rmd^D
x_n\phi_0(x_1)...\phi_0(x_n) \Pi_n(x_1,...,x_n)\,. 
\ee
Up to 4-pt functions, the correlation functions that appear in
(\ref{Wphi0}) are given 
explicitly in (\ref{Pi2})-(\ref{Pi4}) of Appendix A. 
The Legendre transform (\ref{Legendre}) of (\ref{Wphi0}) may be
written as
\be
\label{JA}
J[A] = \sum_{n=0}^{\infty}\frac{1}{n!}\int\rmd^D x_1...\rmd^D
x_n A(x_1)...A(x_n) P_n(x_1,...,x_n)\,. 
\ee
Up to 4-pt functions, the $P$-functions in (\ref{JA}) are related to
the $\Pi$-functions in 
(\ref{Wphi0}) as shown in (\ref{pi2p2})-(\ref{pi4p4}) of Appendix
A. 

Now the requirement that the $P$-correlation functions in
(\ref{JA}) are correlation functions of the free UV $O(N)$ vector
model comes into play. This provides the necessary dynamical principle
for the evaluation of the $P$-correlation functions. At this point we need
to make an assumption for the normalization of the 2-pt functions of
the elementary $N$-component scalars of the model. We can choose to
represent them as unit normalized 2-pt functions of free massless
scalars in $d$ dimensions, i.e. 
\be
\label{norm2pt}
\langle\phi^a(x_1)\phi^b(x_2)\rangle =
\frac{\delta^{ab}}{(x_{12}^2)^{\frac{d}{2}-1}}\,,\,\,\,\,
a=1,2,..,N\,.
\ee
Notice that if $\tilde{\cal O}(x)$, whose
correlation functions are given in (\ref{JA}), is required to be
proportional to 
$\phi^2(x)$, the two parameters $d$ and $D$ are related as
\be
\label{Dd}
D-3 = 2(d-3)\,.
\ee
We are going to set $D=d=3$ at the end, but one may view
(\ref{Dd}) as the relation between two different 
regularization methods based on the analytic continuation in the
number of spacetime dimensions around 3.
Using (\ref{Pi2}) and (\ref{pi2p2}), the relation between
$\tilde{\cal O}(x)$ and $\phi^2(x)$ can be found to be
\be
\label{tildeO}
\tilde{{\cal O}}(x) \equiv
\frac{k}{\sqrt{N}}\phi^a(x)\phi^a(x)\,,\,\,\,\, k^2 =
\frac{\Gamma\left(\frac{D-1}{2}\right)}{4\pi^{\frac{D+1}{2}}}\,.
\ee
Next, using
(\ref{tildeO}) and (\ref{norm2pt}) we obtain 
\be
\label{freeP3}
P_3(x_1,x_3,x_4) \equiv
\frac{8k^3}{\sqrt{N}}
\frac{1}{(x_{12}^2x_{13}^2 
x_{23}^2)^{\frac{D-1}{4}}}\,.
\ee
Finally, from the relation
\be
\label{Pi3P3}
\Pi_3(x_1,x_2,x_3) = \left[P_3(x_1,x_2,x_3)\right]^{amp.}\,,
\ee
and using the D'EPP formula (\ref{depp}) to amputate by $[P_2]^{-1}$ the
free 3pt function (\ref{freeP3}) we obtain
after some algebra
\be
\label{g3}
g_3^2
=\frac{1}{N}\frac{1}{2\kappa_{D+1}^2}\frac{2^4(D-3)^2\pi^{\frac{D+3}{2}}
  \left[\Gamma\left(
    \frac{D-1}{2}\right)\right]^3}{\left[\Gamma\left(\frac{D-1}{4}\right)\right]^6 
 \left[\Gamma\left( \frac{D+3}{4}\right) \right]^2}
\ee
The result (\ref{g3}) is consistent with the results of
\cite{tassosON1} where is was found that the 3-pt function of the
scalar field with 
dimension $2+O(1/N)$ vanishes at the interacting fixed point of the
three dimensional $O(N)$ vector model. Moreover, (\ref{g3}) shows that
for $D=3$ the cubic coupling in the
AdS action (\ref{adsaction}) vanishes. This result is independent of
whether or not the bulk Lagrangian (\ref{adsaction}) contains 
higher spins, hence we conclude that the
cubic self interaction scalar coupling of the higher spin AdS$_4$ theory vanishes.

\subsection{The quartic bulk coupling}

To evaluate the quartic bulk coupling using the Legendre
transform it is simpler to set $D=d=3$. 
From (\ref{norm2pt}) and (\ref{tildeO}) we obtain
\be
\label{freeP4}
P_4(x_1,x_2,x_3,x_4) \equiv
\frac{16k^4}{N}\left[
\frac{1}{x_{12}^2 x_{24}^2 x_{43}^2 
x_{31}^2} +\mbox{crossed }\right]\,.
\ee
Then, the calculation we need to perform is
\bea
\label{Pi4P4}
\Pi_4(x_1,x_2,x_3,x_4) &=& \left[P_4(x_1,x_2,x_3,x_4)\right]^{amp.} \nonumber
\\
 &&\hspace{-2cm}+\left\{ \int\rmd^3 x\rmd^3 y
\left[P_3(x_1,x_3,x)\right]^{amp.}\Pi_2(x,y)\left[
  P_3(y,x_3,x_4)\right]^{amp.} +\mbox{crossed}\right\}\,, 
\eea
where the amputation is done with $\left[P_2\right]^{-1}$. 
Evaluating the integrals on the rhs of (\ref{Pi4P4}) and matching the
results with the explicit expression for $\Pi_4$ given in (\ref{Pi4}),
would give the value of the quartic coupling $g_4$. 

Let us start from the rhs of (\ref{Pi4P4}). The integrals have been
calculated in \cite{LR}, for general dimension $D$, in terms of the
invariant ratios 
\be
\label{vu}
u=\frac{x_{12}^2x_{34}^2}{x_{13}^2x_{24}^2} \,,\,\,\,\,\,
v=\frac{x_{12}^2x_{34}^2}{x_{14}^2x_{23}^2}\,,
\ee
(see \cite{LR}, appendix C), although the form of the results does not appear to be
easily manageable. Nonetheless, our purpose here is to find $g_4$ and
for that we only need the leading term in the short distance expansion of
the rhs of (\ref{Pi4P4}) as $u,v\rightarrow 0$. In practice, we can
simplify things further by considering the expansion of the
integrals in terms of the variables $v$ and $Y=1-v/u$ and consider the
leading term in $v$ for $Y=0$.\footnote{This is inspired from OPE studies of
conformal 4-pt functions where the leading term in $v$ with $Y=0$
corresponds to the leading contribution of conformal scalars
\cite{HPR}.}

From the results in Appendix C of \cite{LR} one can see that the
rhs of (\ref{Pi4P4}) has an expansion of the form
\be
\label{RHSP}
\left[\mbox{RHS of (\ref{Pi4P4})} \right] \propto
\frac{1}{(x_{12}^2x_{34}^2)^2}\left( v^2[-A\ln v+B]+...\right)\,,
\ee
where the dots stand for subleading terms. Now, the important point is
that the coefficient $A$ of the $\ln v$ term exactly vanishes. Therefore,
we should not find a leading logarithmic term also in the lhs of
(\ref{Pi4P4}). This condition determines $g_4$.

Before turning to the evaluation of the AdS integrals in $\Pi_4$, we
comment on the vanishing of its leading logarithmic term. In order to
obtain the full 4-pt function of the operator ${\cal O}(x)$ one should
add to $\Pi_4$ the disconnected part. Once this is done, the OPE
analysis of the 4-pt function can be performed. Notice now that the
vanishing of the 3-pt function $\Pi_3$ implies that the field ${\cal
O}(x)$ itself does not appear in the 4-pt function $\langle {\cal
O}(x_1)...{\cal O}(x_4)\rangle$. The next scalar that contributes to
the OPE of this 4-pt function is a scalar field with dimension
$\tilde{\Delta}=4+\tilde{\eta}$ where $\tilde{\eta}=O(1/N)$. Then, the
4-pt function is expected to have the form 
\be
\label{O4pt}
\langle{\cal O}(x_1){\cal O}(x_2){\cal O}(x_3){\cal O}(x_4)\rangle
\sim 
\frac{1}{(x_{12}^2x_{34}^2)^{\Delta}}[1+v^{\frac{\tilde{\Delta}}{2}}
+... ]\,.
\ee
The vanishing of $\ln v$ term in (\ref{RHSP}) implies the
following relation between the anomalous dimension $\eta$ of ${\cal O}(x)$
and $\tilde{\eta}$
\be
\label{anomdim}
\frac{1}{2}\tilde{\eta} -\eta =0\,.
\ee
Given the known values \cite{LR} $\eta=-2^5/3N\pi^2$ and
$\tilde{\eta}=-2^6/3N\pi^2$ we see that (\ref{anomdim}) is satisfied.

Now turn to the calculation of the leading logarithm in the lhs of
(\ref{Pi4P4}). The contribution to this from the AdS-star graph can
be easily found using the general result (\ref{adsstar}) in Appendix B
\be
\label{starlog}
\left. \Pi_4^{star}\right|_{lead.log}=
\frac{g_4}{\pi^6}\frac{2^9}{N} \frac{1}{(x_{12}^2x_{34}^2)^2}v^2\Bigl[-
\frac{1}{6}\ln v+\cdots\Bigl]\,.
\ee
The contribution form the graviton exchanges is more complicated to
evaluate, since the graphs do not reduce into finite sums of conformal
integrals as in the four dimensional case. The direct channel graviton
exchange has been computed in \cite{LeR} for general dimensions, but
here we also need the crossed channels. After some tedious algebra
whose essentials are presented in Appendix B the final result is
\be
\label{gravlog}
\left. \Pi_4^{grav}\right|_{lead.log} = \frac{1}{2\pi^6}\frac{2^9}{N} 
\frac{1}{(x_{12}^2x_{34}^2)^2}v^2[-\ln v]\,.
\ee
Requiring that $\left[\Pi_4^{star} +\Pi_4^{grav}\right]_{lead.log}=0$ we finally obtain
\be
\label{g4}
g_4=-3\,.
\ee

\section{Discussion}

The critical three dimensional $O(N)$ vector model appears to be a very
interesting laboratory for the study of the AdS/CFT
correspondence. In contrast to most other cases of AdS/CFT, here it is
the boundary CFT side that is well understood at strong
coupling. This means that the correlation functions derived from
AdS/CFT coincide with the well-known correlation functions of the
interacting IR fixed point of the $O(N)$ vector model. Moreover, Legendre
transforming the generating functional of the IR fixed point one gets,
to leading order in the $1/N$ expansion, the generating functional for
the free UV fixed point of the $O(N)$ vector model. Notice that the
assumption that the UV and IR generating functionals are related via a
Legendre transform is important dynamical information, in particular
for the IR fixed point. In the present
paper we initiate the evaluation of the AdS dual of the critical
$O(N)$ vector model making use of its connection with the IR fixed
point of the three dimensional CFT. Assuming a minimal form for the
bulk action i.e. without higher spin or derivative couplings, we evaluate the cubic
and quartic self interaction couplings of the bulk scalar that is dual
to the spin-zero ``current'' of the $O(N)$ model. 

The AdS dual of the $O(N)$ vector model is believed to
correspond to a higher 
spin theory. Therefore it should be possible to check our results
for the cubic couplings within the context of the
equations of motion of the minimal bosonic higher spin theory
$hs(4)$. In particular, the vanishing of the cubic 
coupling was 
conjectured in \cite{tassosON1} to indicate a possible underlying discrete
symmetry for the operator ${\cal O}(x)$. In the context of the higher
spin theory this symmetry may be a manifestation that the dual
operator of 
${\cal O}(x)$ might actually be a fermion bilinear.\footnote{I thank
P. Sundell for discussions on this point.} 

Our result (\ref{g4}) for the quartic bulk self interaction coupling
may not be directly applicable to finding the higher spin Lagrangian.
Nevertheless, it is an intermediate result in this direction. For the
full result one would have to take into account the couplings of the
bulk scalar with higher spin fields. Since these couplings are
believed to be fixed \cite{LMR}, by finding the leading logarithm of
the higher spin exchange graphs in AdS$_4$ one should be able to
unambiguously fix the quartic scalar self interaction coupling.
We expect such a calculation to be complicated but straightforward.

Another interesting class of questions that one can ask is the extension of
the bulk-boundary duality to higher orders in $1/N$. At the field
theory side, there 
exist a number of results for the $O(1/N)$ corrections to  anomalous
dimensions. These
results should somehow be
reproduced by the bulk theory and this raises the intriguing
possibility that we are dealing here with a quantum gravity theory in AdS$_4$
that yields sensible results. Another important
quantity that has been calculated 
is the $1/N$ correction to $C_T$ in (\ref{CT}) which was found to decrease as one goes from
the UV to the IR fixed points of the $O(N)$ vector model
\cite{tassosON1}. Hence it appears to be a  natural extension of the
$C$-function to odd dimensions and is a measure of the
degrees of freedom at the fixed point. Moreover, on the basis of the
results in \cite{tassosON1}, it was argued in \cite{tassosON2} that
the interacting IR fixed point of the $O(N)$ vector model describes
the symmetry
breaking pattern $O(N)\rightarrow O(N-1)$. For this reason, if the
degrees of freedom coupled to the UV free fixed point are $N$, the
massless degrees of freedom coupled to the interacting fixed point are
$N-1$ \cite{tassosON2}. This raises a puzzle regarding the relation
between the free UV 
and interacting IR fixed points of the $O(N)$ vector model for
subleading $N$. Finally, it is also intriguing that the leading-$N$
free energies at the free UV and interacting IR fixed point of the
$O(N)$ vector model are different and related by a rational factor
4/5 \cite{S,tassos_polyl}. This indicates that a holographic thermodynamical
study of the model may hide interesting surprises. We hope to return to
some of these issues in the near future.  

\subsection*{Acknowledgments}

I would like to thank P. Sundell for interesting discussions and 
J. L. F. Barb\'on for a critical reading of the manuscript.

\begin{appendix}
\appendix

\section{Basic AdS/CFT formulas}

The scalar bulk-to-bulk and bulk-to-boundary propagators of a scalar
corresponding to an operator with dimension $\Delta=D/2+1/2$ that we use
are respectively
\bea
\label{bbulk}
G(x,y) &=&
c_{\Delta}\xi^{-\Delta}{}_2F_1\left(\frac{\Delta}{2}+\frac{1}{2},
\frac{\Delta}{2};\Delta-1;\xi^{-2}\right)\,,\nonumber \\
\label{xi}
\xi^2 &=& \frac{r^2+r'^2+(x-y)^2}{2rr'} \,, \,\,\,\,\,\,
  c_{\Delta} = \frac{\Gamma(\Delta)}{2^{\Delta}\pi^{\Delta}}\,, \\
\hat{K}(r';y,x) &=& C_{\Delta}\left[\frac{r'}{r'^2
  +(y-x)^2} \right]^{\Delta}\,,\,\,\,\, C_{\Delta} =
  \frac{
  \Gamma(\Delta)}{\pi^\frac{D+1}{2}}\,. 
\eea
With the above, the explicit expressions for the $\Pi$-functions in
  (\ref{Wphi0}) are
\bea
\label{Pi2}
\Pi_2(x_1,x_2)&=& C_{\Delta}\frac{1}{ x^{2\Delta}_{12}}\,,\\
\label{Pi3}
\Pi_3(x_1,x_2,x_3)&=&
  -\frac{g_3}{2\pi^D}\sqrt{\frac{2^9}{N}}\frac{\left[\Gamma\left(\frac{\Delta}{2}\right)\right]^3 
  \Gamma\left(\frac{3\Delta}{2}
  -\frac{D}{2}\right)}{\left[\Gamma\left(\frac{1}{2}\right)\right]^3}\,
  \frac{1}{(x^2_{12}
  x^2_{13}x^2_{23})^{\frac{\Delta}{2}}}\,,  \\
\label{Pi4}
\Pi_4(x_1,x_2,x_3,x_4)&=&  -g_4
  C_{\Delta}^4\frac{2^9}{N}\int_{0}^{\infty}\!\!\frac{\rmd  
  r}{r^{D+1}}\int\!\!\rmd^{D}x
  \hat{K}(r;x,x_1) \hat{K}(r;x,x_2)\hat{K}(r;x,x_3)
  \hat{K}(r;x,x_4) \nonumber \\
&&
-\Biggl\{g_3^2C_{\Delta}^4\frac{2^9}{N}\int_{0}^{\infty}\frac{\rmd 
  r\rmd r'}{(rr')^{D+1}}\int\rmd^D x\rmd^D y\Bigl[
  \hat{K}(r;x,x_1) \hat{K}(r;x,x_2)
  G(x,y) \nonumber \\
&&\hspace{2.8cm} \hat{K}(r';y,x_3)
  \hat{K}(r';y,x_4)\Bigl]+(x_2 \leftrightarrow x_3)
  +(x_2 \leftrightarrow x_4) \Biggl\}\nonumber \\
&& + C_{\Delta}^4\frac{2^9}{N}
  \left[\frac{1}{4}I^s_{grav}
  +\frac{1}{4}I^t_{grav}+\frac{1}{4}I_{grav}^u\right]\,.  
\eea

The $s$-channel graviton exchange amplitude can be read from the
results of \cite{D'HFR} in the general form given by
\cite{LeR}. Specializing to $D=3$ and $\Delta=2$ we have
\bea
\label{gravs}
&&I_{grav}^s= \frac{1}{(x_{12}^2x_{13}^2x_{14}^2)^2}
\int\frac{\rmd^3 w\rmd w_0}{w_0^4} f(t') \left\{6\left[\frac{w_0}{w_0^2
+(w-x'_{13})^2}\right]^2 \left[\frac{w_0}{w_0^2
+(w-x'_{14})^2}\right]^2 \right.\nonumber \\
&&\hspace{-1cm}-16\,w_0\,\left(\left[\frac{w_0}{w_0^2
+(w-x'_{13})^2}\right]^3 \!\left[\frac{w_0}{w_0^2
+(w-x'_{14})^2}\right]^2 \!\!\!+\left[\frac{w_0}{w_0^2
+(w-x'_{13})^2}\right]^2 \!\left[\frac{w_0}{w_0^2
+(w-x'_{14})^2}\right]^3\right) \nonumber \\
&&\hspace{3cm}\left. +32\,w_0^2\,\left[\frac{w_0}{w_0^2
+(w-x'_{13})^2}\right]^3 \left[\frac{w_0}{w_0^2
+(w-x'_{14})^2}\right]^3\right\} \,,
\eea
where
\bea
\label{ft}
f(t') &=&
-\frac{\pi}{2}\left[\frac{w_0^2}{(w-x'_{12})^2}\right]^{\frac{1}{2}}
+\frac{2}{3}t'^2
{}_2F_1\left(\frac{1}{2},\frac{3}{2};\frac{5}{2};t'\right)\,, \\
t' &=& \frac{w_0^2}{w_0^2 +(w-x'_{12})^2}\,,\,\,\,\,\,x'^i \,=\, \frac{x^i}{x^2}\,. 
\eea
In (\ref{gravs}), let us for simplicity denote the integrals that
involve the first term in (\ref{ft}) with ${\cal I}^s$ and the ones that
involve the hypergeometric function by $I^s$. Then, in an obvious
notation we write
\be
\label{gravs2}
I_{grav}^s = \frac{1}{(x_{12}^2x_{13}^2x_{14}^2)^2}\left[ 6(I_1^s
+{\cal I}_1^s) -16(I_2^s +{\cal I}_2^s) +32(I_3^s+{\cal I}_3^s)\right]\,.
\ee
The $t$ and $u$-channels are obtained from (\ref{gravs}) by the
interchanges $x'_2\leftrightarrow x'_3$ and $x'_2\leftrightarrow x'_4$
respectively. A graphical representation of the 4-pt function is shown
in Fig.1.
The correlation functions in $W[\phi_0]$ and $J[A]$ are related as
\bea
\label{pi2p2}
\Pi_1(x_1,x_2) &=& -\left[P_2(x_1,x_2)\right]^{-1}\,,\\
\label{pi3p3}
\Pi_3(x_1,x_2,x_3) &=& \left[P_3(x_1,x_2,x_3)\right]^{amp.}\,,\\
\label{pi4p4}
\Pi_4(x_1,x_2,x_3,x_4) &=& \left[P_4(x_1,x_2,x_3,x_4)\right]^{amp.}
\nonumber \\
&&
\hspace{-2cm}-\left\{ \int\rmd^3 x\rmd^3
y\left[P_3(x_1,x_2,x)\right]^{amp.}P_2(x,y)\left[P_3(x_3,x_4,x)\right]^{amp.} 
+\mbox{crossed}\right\}\,.
\eea
The amputation is done with $\left[P_2(x_1,x_2)\right]^{-1}$ and with
the help of the D'EPP formula
\bea
\label{depp}
\int\rmd^D{x}\frac{1}{(x_1-x)^{2a_1}(x_2-x)^{2a_2}(x_3-x)^{2a_3}} =
\frac{U(a_1,a_2,a_2)}{(x_{12}^2)^{\frac{D}{2}
-a_3}(x_{13}^2)^{\frac{D}{2}-a_2} (x_{23}^2)^{\frac{D}{2}-a_1}}\,,
\nonumber \\
U(a_1,a_2,a_3)
=\pi^{\frac{D}{2}}\frac{\Gamma\left(\frac{D}{2}-a_1\right)
\Gamma\left(\frac{D}{2}-a_2\right) 
\Gamma\left(\frac{D}{2}-a_3\right)}{\Gamma(a_1)\Gamma(a_2)\Gamma(a_3)}\,, 
\eea
which is valid for $a_1+a_2+a_3  =D$. 
To obtain the inverse 2-pt function of a scalar field we use the formula
\be
\label{inv2pt}
\left[\frac{1}{x^{2A}}\right]^{-1} = \frac{1}{\pi^D}\frac{
\Gamma\left(D-A\right)
\Gamma(A)}{\Gamma\left(A-\frac{D}{2}\right)
\Gamma\left(\frac{D}{2}-A\right)} \frac{1}{(x^2)^{D-A}}\,.
\ee

%%%%%%%%%%%%%%%%%%%%%%%%%%%%%%%%%%%%%%%%%%%%%%%%%%%%%%%%%%%%%%%%%%%%%%
\begin{figure}[ht]
\centering
\begin{minipage}{11cm}
\centering

\psfrag{G}{$\Pi_4(x_1,x_2,x_3,x_4)\,\,\,\,=$}
\psfrag{A}{$+$}
\psfrag{ax1}{$x_1$}
\psfrag{ax2}{$x_2$}
\psfrag{ax3}{$x_3$}
\psfrag{ax4}{$x_4$}
\psfrag{bx1}{$x_1$}
\psfrag{bx2}{$x_2$}
\psfrag{bx3}{$x_3$}
\psfrag{bx4}{$x_4$}
\psfrag{cx1}{$x_1$}
\psfrag{cx2}{$x_2$}
\psfrag{cx3}{$x_3$}
\psfrag{cx4}{$x_4$}
\psfrag{dx1}{$x_1$}
\psfrag{dx2}{$x_2$}
\psfrag{dx3}{$x_3$}
\psfrag{dx4}{$x_4$}
\psfrag{ex1}{$x_1$}
\psfrag{ex2}{$x_2$}
\psfrag{ex3}{$x_3$}
\psfrag{ex4}{$x_4$}
\psfrag{fx1}{$x_1$}
\psfrag{fx2}{$x_2$}
\psfrag{fx3}{$x_3$}
\psfrag{fx4}{$x_4$}
\psfrag{gx1}{$x_1$}
\psfrag{gx2}{$x_2$}
\psfrag{gx3}{$x_3$}
\psfrag{gx4}{$x_4$}

\includegraphics[width=11cm]{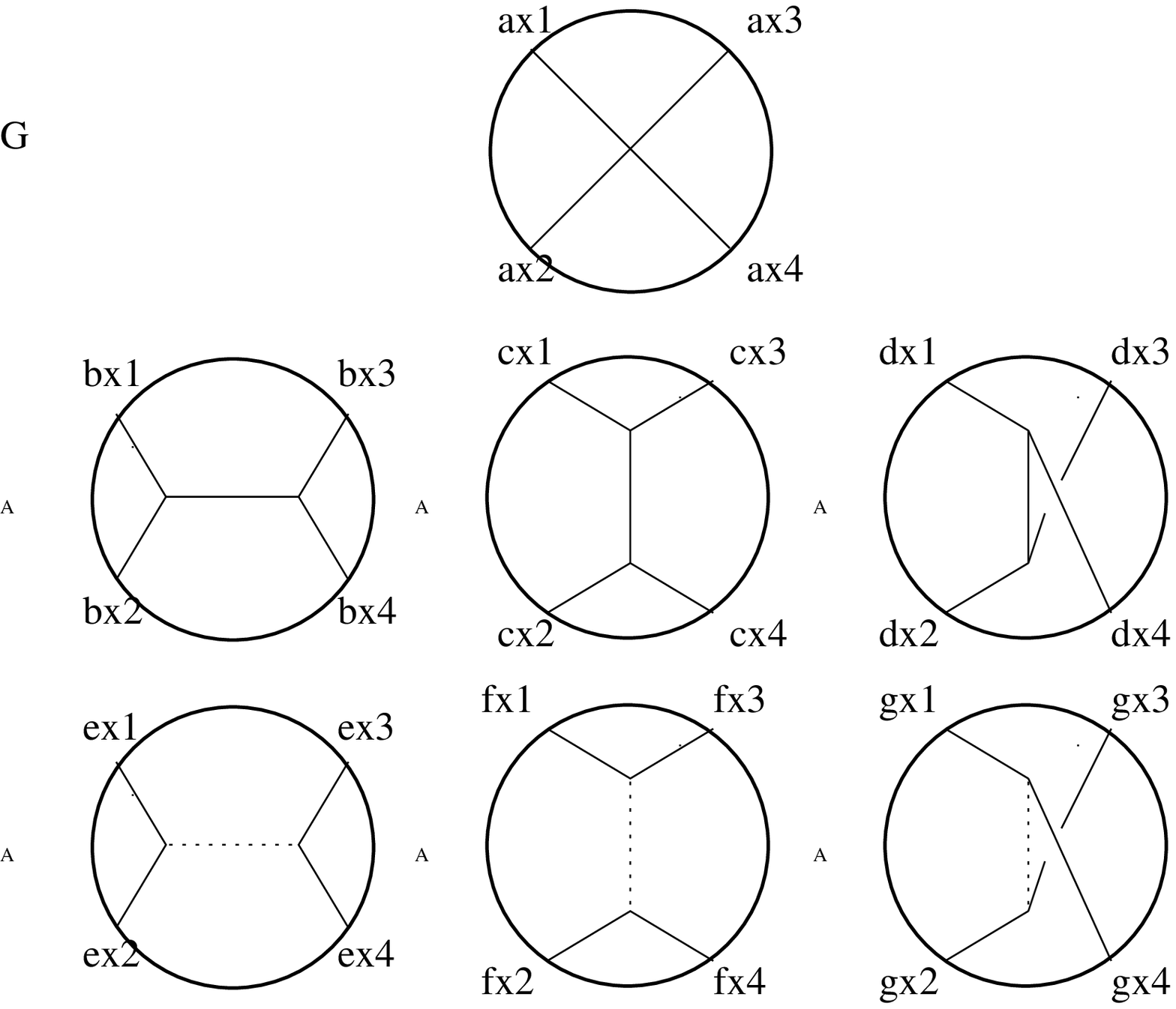}
\centering
\caption{\it \small Graphical representation of
$\Pi_4(x_1,x_2,x_3,x_4)$. The solid lines correspond to the scalar and
the dotted lines to the graviton.}
\end{minipage}
\end{figure} 
%%%%%%%%%%%%%%%%%%%%%%

\section{Leading logarithmic singularities of AdS integrals}

Recall the definition of the cross ratios involved in the calculation
of conformal 4-pt functions
\be
\label{crossr2}
u=\frac{x_{12}^2x_{34}^2}{x_{13}^2x_{24}^2} \,,\,\,\,
v=\frac{x_{12}^2x_{34}^2}{x_{14}^2x_{23}^2}\,,\,\,\, Y=1-\frac{v}{u}\,.
\ee
Then, the general conformal 4-pt function can be expanded in the
variables $v$ and $Y$ which makes it easier to read the contributions due
to various conformal tensors. For example the leading (in the limit
$x_{12}^2\,,\,x_{34}^2\rightarrow 0$), contribution due to a tensor of
dimension $\Delta$ and rank $k$ is of the form \cite{HPR}
\be
\label{rankk}
v^{\frac{\Delta-k}{2}}Y^k\,.
\ee
Using standard techniques for the calculation of AdS
graphs we can easily evaluate the first term on the rhs of (\ref{Pi4})
that corresponds to the AdS-star graph in Fig.1. We give for
completeness the result
\bea
\label{adsstar}
\frac{g_4}{\pi^6}\frac{2^9}{N}
\frac{1}{(x_{12}^2x_{34}^2)^2}v^2\Biggl\{\sum_{n,m=0}^{\infty}
\frac{v^nY^m}{n!m!}
\frac{\Gamma^2(2+n)\Gamma^2(2+2n+m)}{\Gamma(1+n)\Gamma(4+n+m)}
\left[-\ln v+2\psi(4+2n+m)\right. \nonumber \\ \left. 
+2\psi(1+n)-2\psi(2+n)-2\psi(2+n+m)\right]\Biggl\}\,.
\eea
To calculate the leading logarithm of the graviton exchange graph we
start with the integrals in (\ref{gravs}) that come from the
hypergeometric function in (\ref{ft}). For clarity we present here the
integral in the first line of (\ref{gravs}). Using the following
representation \cite{Gradsteyn}
\be
\label{hyper1}
{}_2F_1(a,b;c;z) =\frac{\Gamma(c)}{\Gamma(a)\Gamma(b)}\frac{1}{2\pi{\rm
i}} \int_{\cal C} \rmd s
\Gamma(-s)\frac{\Gamma(a+s)\Gamma(b+s)}{\Gamma(c+s)} (-z)^s\,,
\ee
with an appropriately chosen contour ${\cal C}$ parallel to the
imaginary axis and a standard Feynman parametrization we obtain
\bea
\label{I1}
I^s_1 &=& 
\frac{\pi^2}{4}\frac{1}{2\pi{\rm i}}\int_{\cal C}\rmd s
\Gamma(-s)\Gamma(1+s)
\tilde{I}^s_1\,,\\
\label{tildeI1}
\tilde{I}_1^s &=& \int_0^\infty \!\!\!\!\rmd t_1..\rmd t_3 \,t_1^{1+s}\,t_2t_3
\left(\sum t\right)^{-4-s} \!\!\!\exp\left[\frac{-1}{\sum t}\left[t_1t_2 A_1
+t_1+t+3 A_2 +t_2t_3 A_3\right]\right]\!\!,\\
\label{As}
\sum t &=& t_1+t_2+t_3\,,\,\,\,\,
A_1=\frac{x_{23}^2}{x_{12}^2x_{13}^2}\,\,\,\,
A_2=\frac{x_{24}^2}{x_{12}^2x_{14}^2} \,\,\,\,
A_1=\frac{x_{34}^2}{x_{13}^2x_{14}^2} \,. 
\eea
Next we set in (\ref{tildeI1})
\be
\label{ttoa}
t_3 = \a_1\a_2\a_3\,,\,\,\,\, t_2=\a_1\a_2(1-\a_3)\,,\,\,\,\,
t_3=\a_1(1-\a_2) \,\,\, 0\leq \a_1<\infty\,, 0\leq \a_2,\a_3\leq 1\,,
\ee
and do successively the $\a_1$ and $\a_2$ integrations with result
\be
\label{tildeI1a3}
\tilde{I}_1^s = A_1^{-2}B\left(2,2+s\right)
\int_0^1\rmd\a_3(1-\a_3)\left[1-\a_3 Y\right]^{-2}
{}2F_1\left(2,2;4+s;1-\frac{\a_3(1-\a_3)v}{1-\a_3 Y}\right)\,.
\ee
To obtain the leading logarithm now is suffices to set $Y=0$ in
(\ref{tildeI1a3}). Then, we may use the following representation for the
hypergeometric function \cite{Gradsteyn}
\bea
\label{hyper2}
\frac{1}{2\pi{\rm i}}\int_{{\cal C}'} \rmd t
\Gamma(-t)\Gamma(c-a-b-t)\Gamma(a+t)\Gamma(b+t) (1-z)^t = \nonumber \\
\hspace{1cm}= \Gamma(c-a)
\Gamma(c-b)\frac{\Gamma(a)\Gamma(b)}{\Gamma(c)}{}_2F_1(a,b;c;z)\,,
\eea
where ${\cal C}'$ run parallel to the imaginary axis, and we do the
$\a_3$ integration to end up with a double Mellin-Barnes 
integral over $t$ and 
$s$. The $t$ integration is straightforward while there are double
poles in the $s$ integration. These are handled with the help of the
general formula \cite{HPR}
\be
\label{dpoles}
\frac{1}{2\pi{\rm i}}\int_{\cal C}\rmd s \Gamma^2(-s) g(s) v^s =
\sum_{n=0}^\infty \frac{v^n}{(n!)^2}\left[2\psi(1+n)g(n) -g(n)\ln v
-\frac{\rmd}{\rmd \xi}[g(\xi)]_{\xi=n}\right]\,.
\ee
Keeping only the leading term in $v$ we obtain
\be
\label{I1fin}
\left.I_1^s\right|_{lead. log} =
\left(\frac{x_{12}^2x_{13}^2}{x_{23}^2}\right)^2[-\ln 
v] \frac{\pi^2}{24}\,. 
\ee
Following the same procedure we can find the leading logarithmic terms
in the direct channel as
\bea
\label{I2fin}
\left.I_2^s\right|_{lead. log} &=&
\left(\frac{x_{12}^2x_{13}^2}{x_{23}^2}\right)^2[-\ln 
v] \frac{5\pi^2}{96}\,,\\
\label{I3fin}
\left.I_2^s\right|_{lead. log} &=&
\left(\frac{x_{12}^2x_{13}^2}{x_{23}^2}\right)^2[-\ln 
v] \frac{7\pi^2}{384}\,.
\eea
It is also easy to see, either by direct calculation or from the
results of \cite{LR} that the ${\cal I}^s$ integrals do not have any
logarithmic terms. Then, from (\ref{I1fin})-(\ref{I3fin}) and
(\ref{gravs2}) we see that the leading logarithmic contribution in the direct channel
vanishes.

In general, the calculation of the crossed $t$ and $u$ channels is considerably more
complicated, but the extraction of the leading logarithms can be done
relatively easy on the lines sketched above. For the $I^t$ and $I^u$
integrals we find that their leading logarithms are exactly the same
as the ones of the corresponding $I^s$ integrals. Therefore, their
contribution to the leading logarithm of the graviton exchange graph
vanishes. Hence, the only possible logarithms can come from the
crossed channel integrals ${\cal I}^t$ and ${\cal I}^t$.
Our calculation, done on the lines described above, yield for the
terms that give a non vanishing contribution
\bea
\label{calIt}
\left.{\cal I}^t_1\right|_{lead.log} = \left.{\cal I}^u_1\right|_{lead.log}
=-\left(\frac{x_{12}^2x_{13}^2}{x_{23}^2}\right)^2[-\ln  
v] \frac{\pi^2}{2}\,, \\
\label{calIu}
\left.{\cal
  I}^t_2\right|_{lead.log}=\left.{\cal I}^u_2\right|_{lead.log} 
=-\left(\frac{x_{12}^2x_{13}^2}{x_{23}^2}\right)^2[-\ln  
v] \frac{\pi^2}{4}\,.   
\eea
Using these result we obtain (\ref{gravlog}) in the main text.

\end{appendix}

\end{document}